\documentclass[10pt,a4paper]{article}
\usepackage[utf8x]{inputenc}
\usepackage[left=2cm, right=2cm, top=1.50cm, bottom=1.50cm]{geometry}
\usepackage{amsmath}
\usepackage{amsfonts}
\usepackage{amssymb}
\usepackage{graphicx}
\usepackage{theorem}
\usepackage{subfig}
\usepackage{epstopdf}
\usepackage[hyphens]{url}
\newcommand{\Hi}{$\mathcal{H}_{\infty}$}
\usepackage{url}
\theorembodyfont{ \normalfont }
\newtheorem{theorem}{\it Theorem}
\usepackage{hyperref}

\begin{document}
\begin{flushleft}
Proceedings 	of the {\em Oitavo Encontro dos Alunos e Docentes do Departamento de Engenharia de Computa\c{c}\~{a}o Automa\c{c}\~{a}o At: campinas SP Brazi, 2015}, available in   \href{https://www.fee.unicamp.br/sites/default/files/departamentos/dca/eadca/eadcaviii/artigos/sessao6/olate_et_al.pdf}{[$2015$ EADCA-VIII]}.
\end{flushleft}
\vspace{6 mm}

\begin{center}
\textbf{	{\LARGE   An Approach to Energy Efficiency in a Multi-Hop Network Control System through a Trade-Off between \Hi~Norm and Global Number of Transmissions} } 
\end{center}
\vspace{3 mm}		
\textbf{Jonathan M. Palma, Leonardo de P. Carvalho and Alim P. C. Gon\c{c}alves (Orientador)}

		Departamento de Engenharia de Computa\c{c}\~{a}o e Automa\c{c}\~{a}o Industrial (DCA).    Faculdade de Engenharia El\'{e}trica e de Computa\c{c}\~{a}o (FEEC).  Universidade Estadual de Campinas (Unicamp).   		Caixa Postal 6101, 13083-970 -- Campinas, SP, Brasil. 	{\tt  email \{jmpalma,lcarvalho,alimped\}@dca.fee.unicamp.br}\\

\textbf{ \em Abstract} \textbf{-- The present work proposes a new approach to energy efficiency for filtering of a system interconnected by a Multi-Hop network. The minimization of the energy cost per time unit is obtained through limiting the number of packets retransmission in the Hop-by-Hop mechanism. We explore a trade-off between system performance, measured by estimation error \Hi~norm and energy consumption. The proposal is validated using a theoretical model for energy consumption of MICA2 transceivers.}\\

\textbf{{\em keywords:  MJLS, Hop-by-Hop, Energy-Efficiency, NCS}}

	\section{Introduction}\label{sec:introducao}
	The classical design of a control application immersed in a digital network requires high service quality, ensuring a full-reliable communication and  null or low delay. The information transported through a network has an operational cost and a probability of failure, we usually  minimize the probability of failure by increasing  the number of transmitted packets. This approach increases power consumption and delay  per package.  Using semi-reliable communication in  control system  implies in the performance degradation which sometimes can lead to instability.
	
	Another application that requires a full-reliable communication is the image transmission, where the packet loss can cause useful information  loss in an image \cite{Durans}. The image transmission in particular wireless technologies as Low Rate Wireless Personal Area Network (LR-WPANs) can be complex  where image reconstruction does not admit information loss, causing high energy consumption for the network.
	Wireless sensor network (WSN), a type of LP-WPAN, implemented with vision modules is useful for a wide range of applications. A new field of study is developed to find ways to minimize the implementation cost in image transmission by using semi-reliable communication reconstructing the lost packets using techniques like data selectivity, packet correlation, and others \cite{costa2012discrete,duran}. The use of a semi reliable network instead of a full reliable one has a direct impact in the energy cost due to the decrease in the expected number of transmissions and receptions per packet.
	
\begin{figure}[th!]
		\begin{center}
			\includegraphics[width=1\columnwidth]{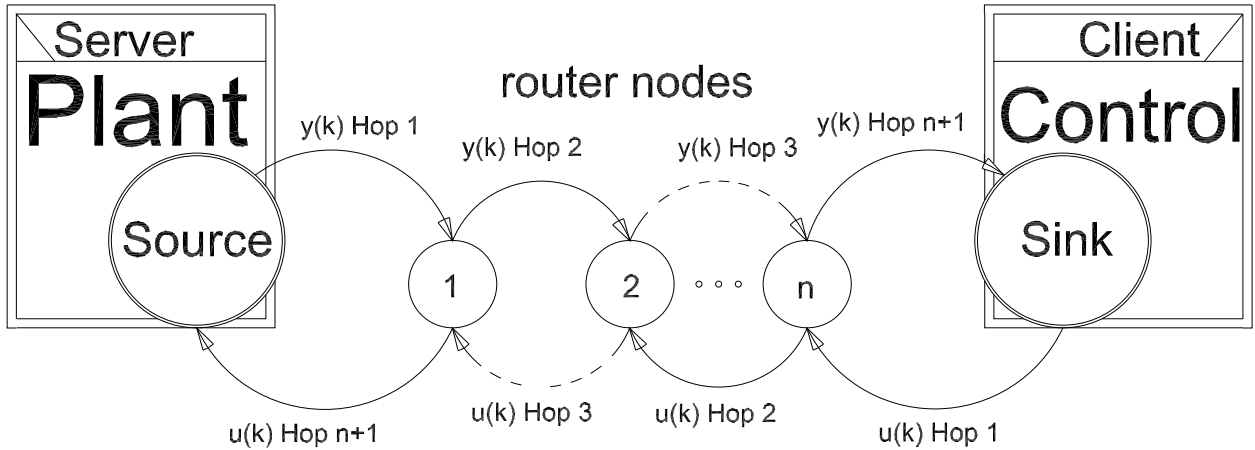}
			\caption{Control  in  Multi-Hop network.}
			\label{modelo}
		\end{center}
	\end{figure}

	The reference \cite{ecu2015} proposes an approach similar to the problem with image transmission in LR-WPAN. This approach allows us to use a Multi-Hop network in filtering applications by decreasing the dynamic system performance in order to minimize the expected number of global transmissions. This techniques propose a trade off between the average number of transmissions and the estimation error \Hi~ norm. In the present work we propose a new approach for energy efficiency of a Network Control System (NCS). We use the theoretical energy consumption model for the MICA2 transceivers \cite{mica} and apply it in the network using a Hop by Hop transport scheme limiting the maximum number of retransmissions per link. Figure \ref{modelo}  represents  a control scheme in a Multi-Hop network as described by \cite{Wang:2008:NCS:1481568}.
	\section{Preliminary}
	\subsection{Markovian Jump Linear System (MJLS)}  
	Consider the following discrete-time system,
	\begin{equation}\label{eq:sistema}
 \mathcal{G}:
		\begin{cases}
		x(k+1)&=A(\theta_k)x(k) + J(\theta_k)w(k),\\
		y(k)&=C_y(\theta_k)x(k) + E_y(\theta_k)w(k),\\
		z(k)&=C_z(\theta_k)x(k) + E_z(\theta_k)w(k),
		\end{cases}
	\end{equation}
	where $x(k) \in \mathbb{R}^n$ is the state vector and $w(k) \in \mathbb{R}^m$ is the external disturbance. The vector $z(k) \in \mathbb{R}^r$ contains the signals to be estimated by the filter through the measured outputs $y(k) \in \mathbb{R}^q$. The random variable $\theta_k \in \mathbb{K}=\{1,2 \ldots, N_n \}$ assumes its values according to a Markov Chain with probability matrix $\mathbb{P}=[p_{ij}]$, where $p_{ij}={\rm P}(\theta_{k+1}=j | \theta_k =i)$. Clearly, $p_{ij}$ are always positive and the sum of the elements in a row of $\mathbb{P}$ is equal to one. To simplify the notation, whenever $\theta_k=i$, we write $A(\theta_k)=A_i$, $J(\theta_k)=J_i$ and so forth.
	
	An important assumption for filter and control design for MJLS is the availability of the Markov mode $\theta_k=i \in \mathbb{K}$ to the controller or filter. If the mode is available we say the design is mode-dependent, or mode-independent otherwise. An intermediate approach considers partial availability, that is the mode itself is not available but only the cluster $U_\ell$ it belongs to. For more details refer to \cite{do2002h}. A particular case of the Markov chain is the generalized Bernoulli jump chain where $p_{ij}=p_{j}$ for all $i,j \in \mathbb{K}$, that is the probability matrix has identical rows. In the following sections  we will consider such assumption.
	\subsubsection*{Optimum \Hi~ filter design for a MJLS}
	Among the markovian filters described in \cite{gonccalves2009controle}, we use a particular filter, the mode-independent one, to minimize estimation error \Hi~ norm. A Luenberger filter that depends on the cluster availability of the Markov mode $\theta_k \in \mathbb{U}_\ell$ is given by
	\begin{equation}\label{eq;filtro}
 \mathcal{O}:
		\begin{cases}
		x_f(k+1)&= A_{\ell}x_f(k) + B_{f\ell}(y(k)-C_{y\ell}x_f(k)),\\
		z_f(k)&= C_{z\ell}x_f(k)+D_{f\ell}(y(k)-C_{y\ell}x_f(k)).
		\end{cases}
	\end{equation}
	Note that, we are assuming that $A_i=A_\ell$, $C_{yi}=C_{y\ell}$ and $C_{zi}=C_{z\ell}$ for all $i \in \mathbb{U}_\ell$, that is, the plant matrices do not vary within a given cluster. Notice that mode-independent design is clearly a particular case for cluster design with only one cluster containing all Markov chain modes.
	\begin{theorem}\label{thm1}
		There exists a filter of the form (\ref{eq;filtro}) such that $ {\|\mathcal{G}_o \|^2_\infty} < {\gamma}$ if and only if there exist symmetric matrices $H_i$, $X$, and the matrices $F_\ell$, $K_\ell$ with compatible dimensions that satisfy the LMIs
		\begin{equation}\label{eq:LMI1}
		\begin{bmatrix}
		H_i & \bullet & \bullet & \bullet \\
		0 & \gamma I & \bullet & \bullet\\
		XA_\ell+F_\ell C_{y\ell} & XJ_i+F_\ell E_{yi} & X & \bullet\\
		C_{z\ell}-K_\ell C_{y\ell} & E_{zi}-K_\ell E_{yi} & 0 & I \\
		\end{bmatrix} > 0,\end{equation}
		\begin{equation}\label{eq:LMI2}
		\sum_{j=1}^{N_n} p_{j}H_j- X<0,
		\end{equation}
		for all $i \in \mathbb{U}_\ell \cap \mathbb{K}$ and $\ell \in \mathbb{L}$. The filter gains are given by, $ \ B_{f\ell}=-X^{-1}F_\ell, \ \ D_{f\ell}=K_\ell$.
		The proof of this theorem can be found in \cite{fioravanti2015optimal}.
	\end{theorem}
	\subsection{Hop by Hop transport model}
	The Hop-by-Hop is a  mechanism  that distributes transport control along the Source-Sink path commonly used in wireless networks, where acknowledgement is transmitted locally  between each hop. The  Hop-by-Hop model and notation can be seen in \cite{chicohop,heimlicher2007end}. The probability of successful hop-by-hop transmissions $P_S$ is determined according to,
	\begin{equation}
	P_S=[1-(1-p)^L]^N.\label{ps}
	\end{equation}
	where $N$ is the number of hops, $L$ is the maximum allowed number of transmissions, $p$ is the link data delivery probability. This probability is directly related to the Markov chain governing the MJLS presented in the previous section.
	\subsubsection{Fundamental Concepts for Energy-Consumption}
\begin{figure}[h!]
		\begin{center}
			\includegraphics[width=0.73\columnwidth]{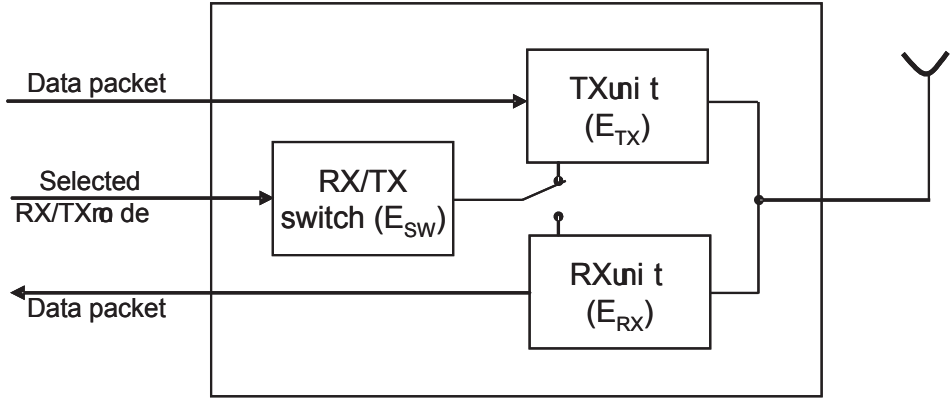}
			\caption{Energy model of radio transceiver.}
			\label{ec}
		\end{center}
	\end{figure}

	A network is composed by nodes, each one has a transceptor unit which can work in two modes: a transmission mode ($TX$) and a reception mode ($RX$). It is possible to define  the energy consumption: $E_{SW}$, $E_{TX}(Lp,P_{out})$ and $E_{RX}(Lp)$, which are the energies used to switch between modes, to transmit for power $ P_{out}$  and to receive a packet, respectively. Both $TX$ and $RX$ energy consumptions depend on the  packet length $Lp$. Figure \ref{ec} corresponds to the schematized energy consumption per transceptor \cite{lecuire2008energy}.
	In  control system applications, the packets with the measurements are transmitted in constant intervals determined by the sampling rate. The energy consumed by time per unit in a multi-hop network corresponds to the mathematical expectation of the energy consumed per package $\mathcal{E}$ divided by the time between samples $Ts$. The theoretical models of energy consumption may consider more elements, which can be seen in \cite{lecuire2008energy,costa2012discrete}.
	\section{Numeric example}
\begin{figure}[th!]
		\centering
		\subfloat[Plant]{\includegraphics[height=0.30\columnwidth]{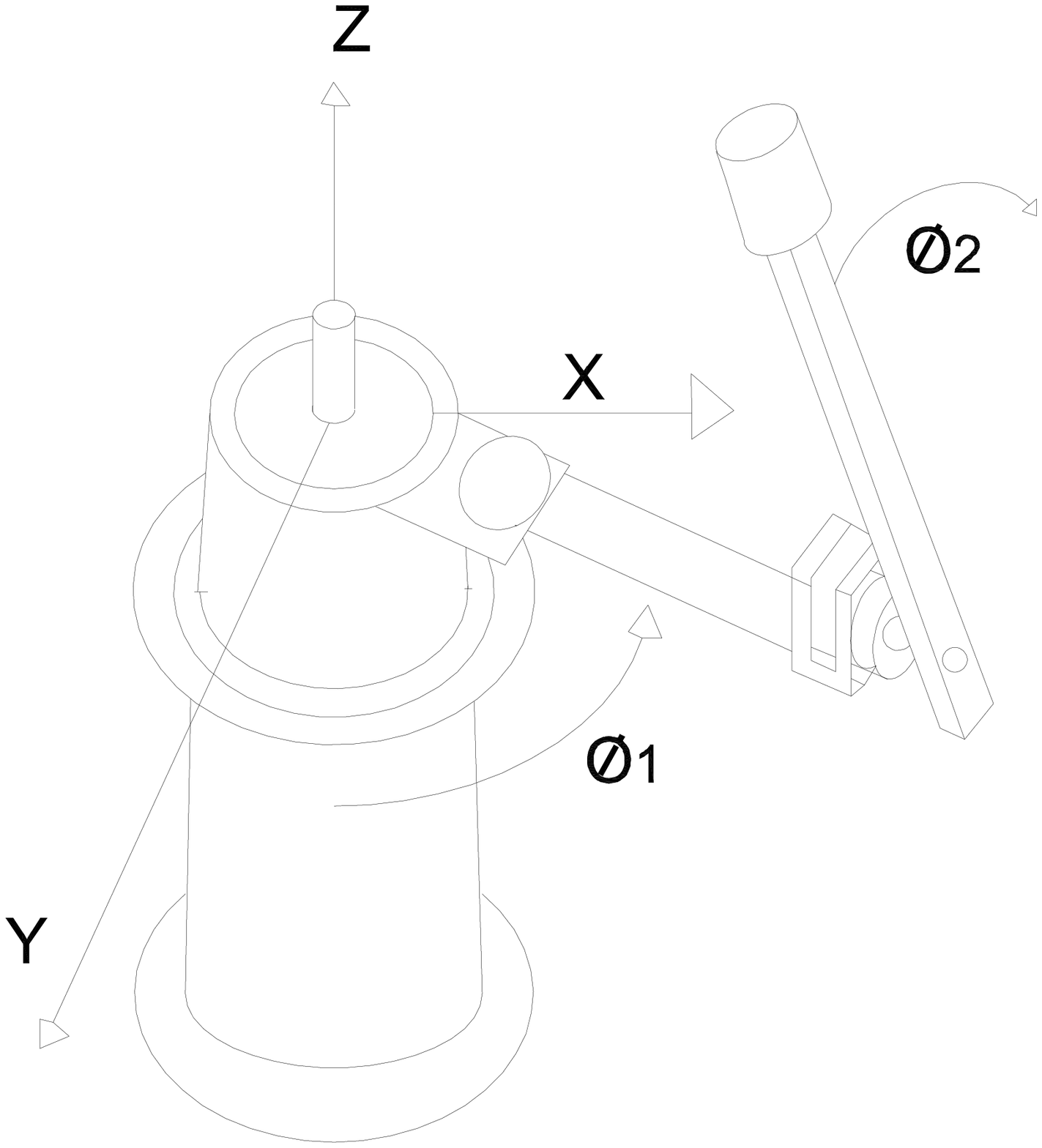}\label{p03}}
		\subfloat[ Norm]{\includegraphics[height=0.3\columnwidth]{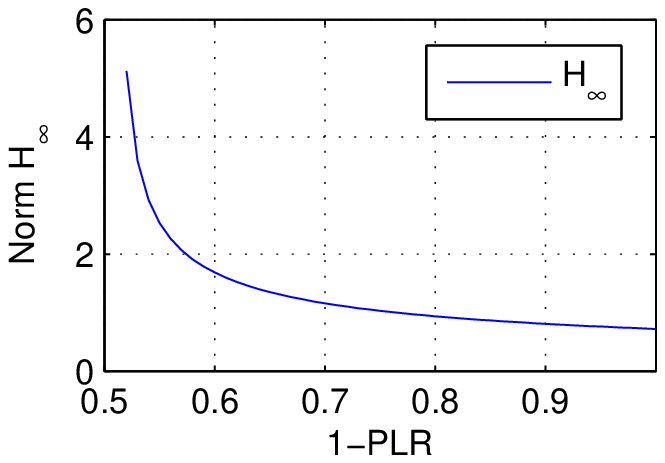}\label{p04}}
		\label{ss2}
	\end{figure}

	The plant used in the \cite{ecu2015} is a rotary inverted pendulum \cite{andre}, in which the joint angle is measured and transmitted every $50$[ms] and we used a zero-order holder with a sampling frequency of $50$[ms] to obtain a discrete-time model. The project is to design a state observer filter in semi-reliable communication using the Hop by Hop scheme and limiting the maximum allowed retransmissions per packet $L$. We consider a mode-independent filter, that is, $N_c=1$. Figure \ref{p03} is a diagram of the plant and Figure \ref{p04} shows the \Hi~ norm degradation with respect to packet loss rate.
	\subsection{Network Model} The measured signal $y(k)$ is transmitted in a multi-hop network with $10$ router units, whose implemented transport mechanism is the Hop-by-Hop. The energy consumption model is based in the wireless units which operates at $3$V and have the following current consumption: the consumption per transmitted byte is $20$[mA] for a power of $0$[dbm]; for each received byte we spend $15$[mA]; the system also has a fixed current consumption for each transition between modes, $15$[mA/per] switch, the duration of those transitions are $250\times 10^{-6}$[s]. A complete explanation about those values can be found in \cite{lecuire2008energy}. The average energy cost for a network with $10$ routers was obtained by a Monte Carlo simulation with $1\times10^{6}$ iterations per packet. 
	
	\subsection{Analysis Metrics} Through a Monte Carlo simulation we obtained the average energy consumed per unit time for preset values of $L$, $p$ and $N$. In order to better understand the results we created two new metrics, the first one is the rate $\Upsilon_H$ between the norm obtained with packet loss and the one obtained with the classic filtering without loss and the other is the ratio $\Upsilon_E$ between the average energy consumption with a given retransmissions limit and the average without any limit defined by the equation $\Upsilon_{E}=\frac{\mathcal{E}[E|L<\infty]}{\mathcal{E}[E |L \rightarrow \infty]}$.
	
	\subsection{Preliminary Results}
\begin{figure}[ht!]
		\centering
		\subfloat[$p=0.4$]{\includegraphics[height=0.36\columnwidth]{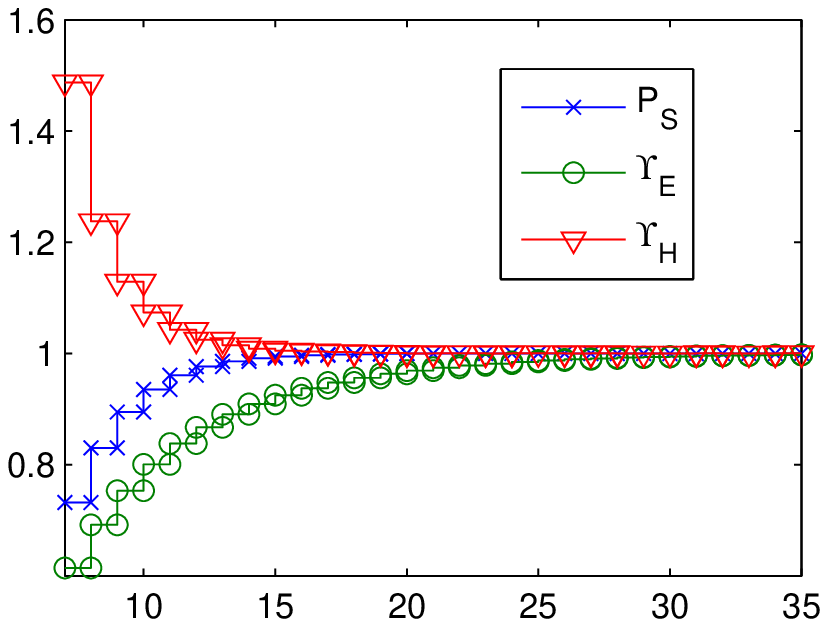}\label{x1}}
		\subfloat[$p=0.5$]{\includegraphics[height=0.36\columnwidth]{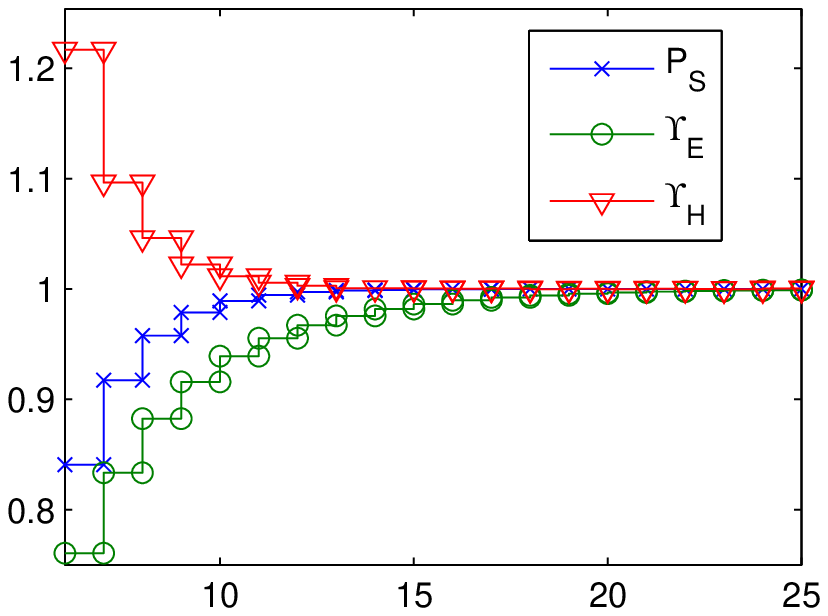}\label{x2}}
		
		\subfloat[$p=0.6$]{\includegraphics[height=0.36\columnwidth]{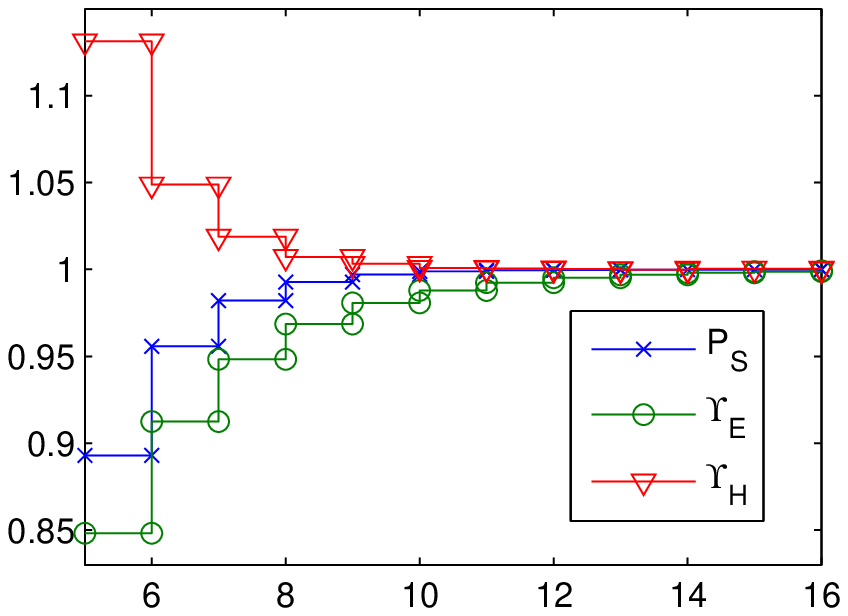}\label{x3}}
		\subfloat[$p=0.7$]{\includegraphics[height=0.36\columnwidth]{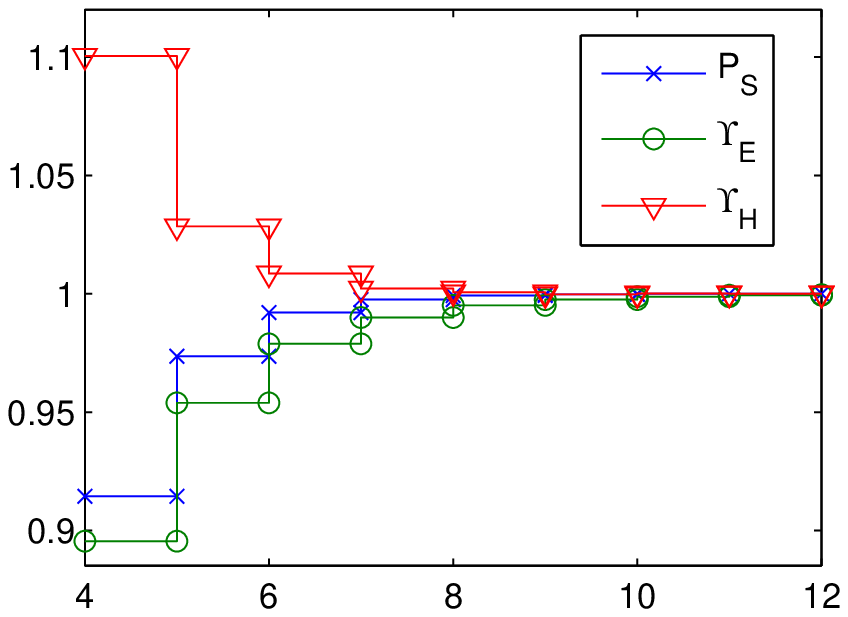} \label{x4}}
		\caption{Variation the metric by  $L$.}
		\label{xx}
	\end{figure}

	This section displays the results in the same way as \cite{ecu2015}. For our model of energy consumption, the \Hi~norm converges faster to the classical than the energy consumption per time unit. The packet length used was 25 bytes contemplating the header. Figure \ref{xx} plots the metrics obtained for four values of $p$ in function of $L$.
	Through simulations it was possible to observe that, using the energy model proposed in this work, the energy consumption behavior has a direct correlation with the number of transmissions due the cost $E_{SW}$ in function of the number of transitions where the units found in reception mode and if the unit needs to transmit it switches to transmission mode and then go back to reception mode. Due the results of Monte Carlo simulation we obtained the average number of receptions have $p$ times the number of transmissions, where $p\mathcal{E}(TX)\approx \mathcal{E}(RX)$. 
	\section{Conclusion}
	Limiting the number $L$ in a multi hop network using the Hop by Hop scheme has a direct impact in the decrease of the average energy consumption per packet. The numerical results indicate that the percentage decrease of the average energy consumption is greater than the decrease in performance, as indicated by the \Hi~ norm. Like the image transmission problem  is willing to lose quality in our case the the immunity to disturbs (\Hi) is possible to increase the network life spam. Limit the number of transmission can be useful for the network lowering the congestion and consequently the average delay.
	
	For future work we propose to obtain the complete characterization of the stochastic process which describes the number of $TX$ and $RX$ of the transport model in a Hop by Hop network, in order to use those values in a more complete model of energy consumption. One of the most important perspectives is to formalize the control problem in a Networked Control System, which we term as ``intelligent'', where there is cooperation between the controller and the network. This scheme intends to obtain the value of $L$ that fulfills the performance criteria  given by \Hi~ norm and the life spam, congestion and others. Such a ``intelligent'' Networked Control System would have different controllers for different network status.

	\bibliographystyle{IEEEtran}
	\bibliography{ieee-lars2014}

\begin{thebibliography}{10}
\providecommand{\url}[1]{#1}
\csname url@samestyle\endcsname
\providecommand{\newblock}{\relax}
\providecommand{\bibinfo}[2]{#2}
\providecommand{\BIBentrySTDinterwordspacing}{\spaceskip=0pt\relax}
\providecommand{\BIBentryALTinterwordstretchfactor}{4}
\providecommand{\BIBentryALTinterwordspacing}{\spaceskip=\fontdimen2\font plus
\BIBentryALTinterwordstretchfactor\fontdimen3\font minus
  \fontdimen4\font\relax}
\providecommand{\BIBforeignlanguage}[2]{{%
\expandafter\ifx\csname l@#1\endcsname\relax
\typeout{** WARNING: IEEEtran.bst: No hyphenation pattern has been}%
\typeout{** loaded for the language `#1'. Using the pattern for}%
\typeout{** the default language instead.}%
\else
\language=\csname l@#1\endcsname
\fi
#2}}
\providecommand{\BIBdecl}{\relax}
\BIBdecl

\bibitem{Durans}
C.~Duran-Faundez and V.~Lecuire, ``Error resilient image communication with
  chaotic pixel interleaving for wireless camera sensors,'' in
  \emph{Proceedings of the Workshop on Real-world Wireless Sensor Networks},
  ser. REALWSN 08, 2008, pp. 21--25.

\bibitem{costa2012discrete}
D.~G. Costa and L.~A. Guedes, ``A discrete wavelet transform (dwt)-based
  energy-efficient selective retransmission mechanism for wireless image sensor
  networks,'' \emph{Journal of Sensor and Actuator Networks}, vol.~1, no.~1,
  pp. 3--35, 2012.

\bibitem{duran}
C.~Duran-Faundez, V.~Lecuire, and F.~Lepage, ``Tiny block-size coding for
  energy-efficient image compression and communication in wireless camera
  sensor networks,'' \emph{Signal Processing: Image Communication}, vol.~26,
  no.~8, pp. 466--481, 2011.

\bibitem{ecu2015}
J.~M. Palma, d.~L. Carvalho, A.~Gon{\c{c}}alves, C.~Galarza, and A.~Oliveira,
  ``Application of control theory markov systems to minimize the number of
  transmissions in a multi-hop network,'' in \emph{Proceedings of the
  Asia-Pacific Conference on Computer Aided System Engineering (APCASE)
  conference}.\hskip 1em plus 0.5em minus 0.4em\relax Ecuador: IEEE, Jul. 2015,
  pp. 296--301.

\bibitem{mica}
C.~T. Inc, ``{The Smart Sensor Company},'' \url{http:/moog-crossboe.com}, 2015,
  [Online; accessed 1-July-2015].

\bibitem{Wang:2008:NCS:1481568}
F.-Y. Wang and D.~Liu, \emph{Networked Control Systems: Theory and
  Applications}, 1st~ed.\hskip 1em plus 0.5em minus 0.4em\relax Springer
  Publishing Company, Incorporated, 2008.

\bibitem{do2002h}
J.~B. Do~Val, J.~C. Geromel, and A.~P. Gon{\c{c}}Alves, ``The h 2-control for
  jump linear systems: cluster observations of the markov state,''
  \emph{Automatica}, vol.~38, no.~2, pp. 343--349, 2002.

\bibitem{gonccalves2009controle}
A.~P. d.~C. Gon{\c{c}}alves, ``Controle din{\^a}mico de sa{\'\i}da para
  sistemas discretos com saltos markovianos,'' Ph.D. dissertation, Universidade
  Estadual de Campinas (UNICAMP). Faculdade de Engenharia El{\'e}trica e de
  Computa{\c{c}}{\~a}o, 2009.

\bibitem{fioravanti2015optimal}
A.~Fioravanti, A.~Gon{\c{c}}alves, and J.~Geromel, ``Optimal and
  mode-independent filters for generalised bernoulli jump systems,''
  \emph{International Journal of Systems Science}, vol.~46, no.~3, pp.
  405--417, 2015.

\bibitem{chicohop}
S.~Heimlicher, P.~Nuggehalli, and M.~May, ``End-to-end vs. hop-by-hop
  transport,'' \emph{SIGMETRICS Perform. Eval. Rev.}, vol.~35, no.~3, Dec.
  2007.

\bibitem{heimlicher2007end}
S.~Heimlicher, M.~Karaliopoulos, H.~Levy, and M.~May, ``End-to-end vs.
  hop-by-hop transport under intermittent connectivity,'' p.~20, 2007.

\bibitem{lecuire2008energy}
V.~Lecuire, C.~Duran-Faundez, and N.~Krommenacker, ``Energy-efficient image
  transmission in sensor networks,'' \emph{International Journal of Sensor
  Networks}, vol.~4, no. 1-2, pp. 37--47, 2008.

\bibitem{andre}
A.~M. de~Oliveira, ``Analise e controle de um sistema mecanico com dados
  transmitidos atraves da rede,'' in \emph{Monografia}.\hskip 1em plus 0.5em
  minus 0.4em\relax Unicamp, 2015.

\end{thebibliography}

\end{document}